\documentclass[sigconf]{acmart}

\renewcommand\footnotetextcopyrightpermission[1]{}

\hypersetup{
    colorlinks=true,
    urlcolor=magenta,
    linkcolor=red
}

\usepackage{enumitem}
\usepackage{booktabs}
\usepackage{array}
\usepackage{tabularx}
\usepackage{amsmath}
\usepackage{algorithm}
\usepackage{algorithmicx}
\usepackage{algpseudocode}
\usepackage{xcolor}
\usepackage{multirow}
\usepackage[table]{xcolor}
\usepackage[most]{tcolorbox}

\AtBeginDocument{%
  }

\begin{document}

\title{IntervenSim: Intervention-Aware Social Network Simulation \\ for Opinion Dynamics}

\author{Yunyao Zhang}
\affiliation{%
  \institution{Huazhong University of Science and Technology}
  \city{Wuhan}
  \country{China}
}

\author{Zuocheng Ying}
\affiliation{%
  \institution{Huazhong University of Science and Technology}
  \city{Wuhan}
  \country{China}
}

\author{Xinglang Zhang}
\affiliation{%
  \institution{Huazhong University of Science and Technology}
  \city{Wuhan}
  \country{China}
}

\author{Junqing Yu}
\affiliation{%
  \institution{Huazhong University of Science and Technology}
  \city{Wuhan}
  \country{China}
}

\author{Peng Fang}
\affiliation{%
  \institution{Huazhong University of Science and Technology}
  \city{Wuhan}
  \country{China}
}

\author{Xu Chen}
\affiliation{%
  \institution{Beijing Institute of Computer Technology and Applications}
  \country{China}
}

\author{Wei Yang}
\affiliation{%
  \institution{Huazhong University of Science and Technology}
  \city{Wuhan}
  \country{China}
}

\author{Zikai Song}
\authornote{Corresponding author.}
\affiliation{%
  \institution{Huazhong University of Science and Technology}
  \city{Wuhan}
  \country{China}
}

\renewcommand{\shortauthors}{Yunyao Zhang et al.}

\begin{abstract}
LLM-based social network simulation introduces a new computational approach for modeling event evolution in complex online environments.
However, existing methods typically simulate social processes under a fixed event trajectory, treating the event as static once initialized and overlooking intervention dynamics, and thus fail to capture the intrinsic evolution of real social network events, where source-side interventions and collective interactions continuously reshape event trajectories, sometimes leading to secondary popularity explosions and collective attitude shifts.
To address this limitation, we introduce an \textbf{intervention}-aware \textbf{sim}ulation framework, \textbf{IntervenSim}, that models event evolution and intervention in a closed loop.
We model event developments and source-side interventions using \textit{source agents}, and collective crowd reactions using \textit{crowd agents},
capturing their continuous co-evolution through an \textit{\textbf{intervention-aware mechanism}} that couples source-side intervention, group interaction, and feedback-driven adjustment of subsequent interventions.
Experiments on diverse real-world events show that \textbf{IntervenSim} improves MAPE by 41.6\% and DTW by 66.9\% over prior frameworks, while reducing computational cost with fewer yet more capable agents. These improvements indicate that \textbf{IntervenSim} not only simulates regular event trajectories more faithfully, but also better captures opinion dynamics under intervention in complex cases.
\end{abstract}




\maketitle

\begin{figure*}
    \centering
    \includegraphics[width=\textwidth]{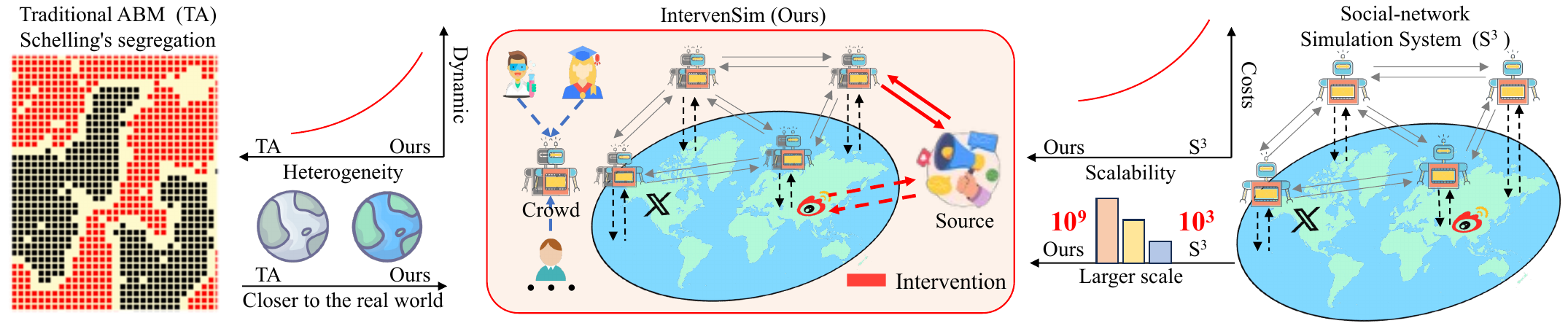}
    \caption{\textbf{Comparison of social simulation paradigms.} 
    \textbf{Traditional ABMs (left)} predict outcomes but lack behavioral heterogeneity and dynamic adaptability. 
    \textbf{LLM-based agents (right)} enhance individual realism but remain limited in scalability and large-scale simulation. 
    \textbf{Our approach (middle)} models society as a dynamic system of crowd and source agents, enabling scalable and realistic simulation of collective behavior.
    }
    \label{fig1:teaser}
\end{figure*}

\section{Introduction}
Simulating how collective behaviors and social structures emerge is central to understanding social diffusion, mobilization, and opinion formation~\cite{Social-Simulation-overview-2014,social-network-analysis2004development}. 
More broadly, social simulation provides a computational lens for studying how individual interactions accumulate into population-level diffusion processes and collective outcomes in open social environments. 
In \textbf{social network simulation (SNS)}~\cite{LLM-Agent-based-simulation-survey-2024large}, human interactions unfold through dynamic exchanges where information propagation, opinion dynamics, and group polarization coevolve across complex relational networks~\cite{diffusion-online-social-networks2017survey,coling-2025-chenxiuying}. 
These interdependent feedbacks make population outcomes highly nonlinear, as small initiating groups can trigger large-scale mobilization and cascading effects~\cite{5rule-lupeng-2018exploring,peak-time-lupeng2018big}. 
Traditional paradigms, including \textit{mechanistic models}~\cite{traditional-simulation-system-dynamics,traditional-simulation-discrete-events}, \textit{empirical and statistical models}~\cite{PSP-2018,peak-height-lupeng-2018predicting,shapes-lupeng2019strength}, and \textit{agent-based models}~\cite{first-agent-based-model-schelling-1971,first-large-scale-agent-model-1996}, capture macro-micro regularities but remain constrained by static parameters and handcrafted behavioral assumptions, limiting their ability to model adaptive cognition, evolving interaction, and intervention-sensitive opinion dynamics in open social events.
These limitations motivate scalable, cognitively grounded frameworks for capturing intervention-aware dynamics and emergent collective behavior~\cite{Micromotives-Macrobehavior-2006,Stanford-town-2023,GAS32025-ga}.

Recent studies show that LLMs~\cite{gpt-4-2023,DeepSeek-OCR2025deepseek,logicagent-2025ambiguity} can endow social agents with reasoning, perception, and interactive capabilities, enabling simulations ranging from small-scale communities~\cite{Stanford-town-2023,Stanford1000agents-2024} to multi-scene social systems~\cite{s3-2023,Socioverse2025socioverse}. 
Advanced frameworks such as \textbf{GA-S\textsuperscript{3}}~\cite{GAS32025-ga}, \textit{MF-LLM}~\cite{MF-LLM2025mf}, \textit{SocioVerse}~\cite{Socioverse2025socioverse}, \textit{AgentSociety}~\cite{Agentsociety-2025agentsociety}, \textit{Yulan-OneSim}~\cite{Yulan-onesim2025-chenxu}, and \textit{OASIS}~\cite{Oasis-2025-shaojing} demonstrate lifelike social interactions; however, they typically operate under fixed event trajectories once initialized, lacking explicit intervention and dynamical evolution modeling, which limits their ability to capture the intrinsic evolution of real social media events.

Accurately modeling social dynamics with complex interventions involves two core challenges. 
\textbf{(C1) Intervention-Aware Modeling}, characterizing how evolving interventions shape group communication and opinion over time, and 
\textbf{(C2) Supporting Requirements for Intervention-Aware Simulation}, which include: 
\textbf{(C2.1) Pre-Intervention Representation}, constructing precise, entity-centric representations of heterogeneous social actors while preserving population-level alignment across diverse social contexts; and
\textbf{(C2.2) Post-Intervention Dynamics}, capturing how group-level attitude shifts and intervention-sensitive updates evolve through ongoing interaction and feedback.

\begin{figure*}[t]
    \centering
    \includegraphics[width=\linewidth, height=0.45\linewidth]{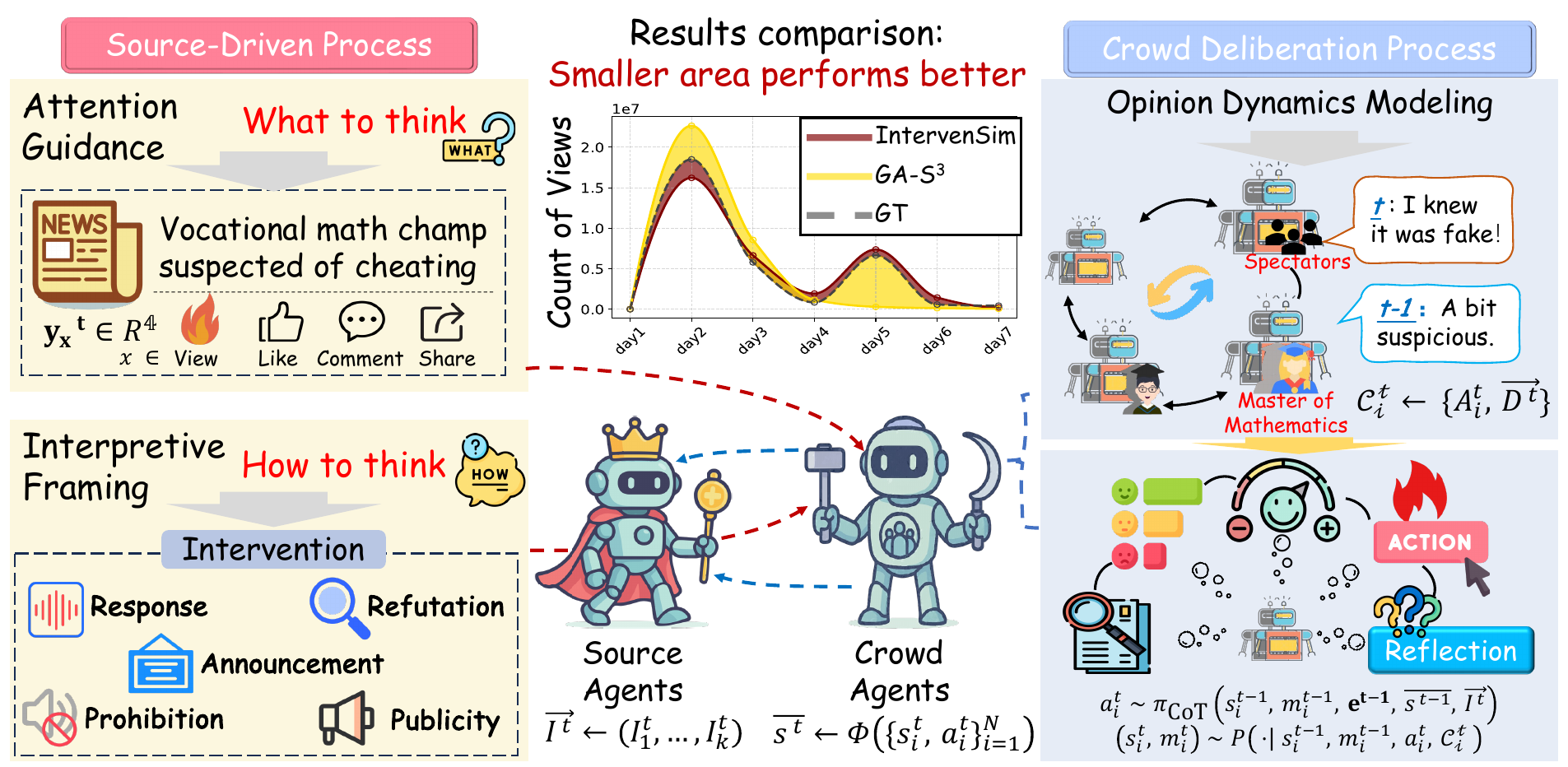}
    \caption{
    \textbf{IntervenSim overview.} The framework comprises two interactive processes that capture top-down intervention and bottom-up public response in social network simulation.
    The \textbf{Source-Driven Process} (\textbf{left}) illustrates how source agents guide public cognition through two coupled stages of intervention: \textbf{Attention Guidance} (\textit{what to think}) and \textbf{Interpretive Framing} (\textit{how to think}), forming a two-stage intervention process that first guides public attention and then shapes interpretation.
    The \textbf{Crowd Deliberation Process} (\textbf{right}) models collective reasoning and opinion evolution through Chain-of-Thought decision making and dynamic attitude updates, illustrating how reflection and interaction shape population-level opinion change.
    The \textbf{comparison} (\textbf{top center}) shows that \textbf{IntervenSim} unifies interpretable reasoning with large-scale social dynamics beyond prior LLM-based simulators.
    }
    \label{fig2:pipline}
\end{figure*}

To address these challenges, we introduce \textbf{IntervenSim} (see Fig.~\ref{fig2:pipline}), an intervention-aware social simulation framework for modeling large-scale event and opinion dynamics under evolving interventions.
Inspired by \textit{agenda-setting theory}~\cite{Theory-Agenda-Setting1972agenda,agenda-melding1999individuals} and classical \textit{opinion dynamics models}~\cite{Theory-degroot-1974reaching,Theory-Hegselmann-Krause-2015opinion}, 
IntervenSim models event evolution and intervention through three tightly coupled components:

\noindent\textit{\textbf{(1) Intervention-Aware Simulation}} (for \textbf{C1}). We formulate event evolution and intervention as a closed-loop simulation process, replacing explicit pairwise agent interactions with structured interactions between \textit{source agents}, which model event developments and source-side interventions, and \textit{crowd agents}, which capture collective responses. Multi-round intervention signaling, group-level interaction, and feedback-driven adjustment jointly coordinate their co-evolution.

\noindent\textbf{(2) \textit{Adaptive Crowd Representation}} (for \textbf{C2.1}). To support intervention-aware simulation, we introduce an adaptive crowd construction strategy based on event relevance and orientation needs, enabling scalable, population-consistent representation of heterogeneous social actors across diverse social contexts. 

\noindent\textbf{(3) \textit{Dynamics Modeling with Intervention Feedback}} (for \textbf{C2.2}). We model how public interactions and source-side interventions jointly shape opinion dynamics over time, capturing emergent outcomes such as consensus and polarization formation. This component further enables intervention-sensitive updates of group states, allowing attitude shifts to unfold through ongoing interaction and feedback.

Experiments on diverse real-world online events show that \textbf{IntervenSim} achieves substantial improvements of \textbf{41.6\%} in MAPE and \textbf{66.9\%} in DTW over prior frameworks, while reducing computational cost with fewer yet more capable agents. 
Case studies further confirm its ability to reproduce emergent phenomena such as intervention-driven attitude shifts and consensus-polarization transitions, showing that \textbf{IntervenSim} not only simulates regular event trajectories in most large-scale ordinary events more faithfully, but also better captures opinion dynamics under intervention in complex cases.

\section{Related Work}

A full discussion of related work appears in Appendix~\ref{appA:related work}; we highlight the most relevant directions here.

\subsection*{Social Simulation Systems}
Traditional social simulation systems model how individual behaviors shape societal structures and diffusion processes~\cite{Social-Simulation-overview-2014,social-network-analysis2004development}. 
They can be grouped into three paradigms. 
\textit{(1) Mechanistic models} describe collective behavior through explicit equations or procedural dynamics such as system-dynamics and discrete-event simulations~\cite{traditional-simulation-system-dynamics,Systems-Based-Approach-2012,traditional-simulation-discrete-events,dynamics-lupeng-2021swarm}. 
They capture macro-level diffusion such as epidemic spreading and online mobilization~\cite{SIR-model2002spread,5rule-lupeng-2018exploring,peak-time-lupeng2018big}, but rely on fixed parameters and lack adaptive cognition. 
\textit{(2) Empirical and statistical models} identify social diffusion patterns from data, including \textbf{PSP}~\cite{PSP-2018} and peak-based participation dynamics~\cite{peak-height-lupeng-2018predicting,shapes-lupeng2019strength}. 
They reveal large-scale regularities, but offer limited insight into the interaction mechanisms underlying opinion dynamics. 
\textit{(3) Agent-based models} capture emergent phenomena through local interactions among heterogeneous agents~\cite{Old-Agent-Based-Social-Simulation-2002,squazzoni2008micro,Multiagent-Systems2005,Multi-Agent-Systems-application-2018}. 
Classical studies such as Schelling's segregation model~\cite{first-agent-based-model-schelling-1971,Cellular-Automata-1998,Micromotives-Macrobehavior-2006} and Sugarscape~\cite{first-large-scale-agent-model-1996} show how simple local preferences produce complex societal patterns. 
Despite these advances, traditional paradigms still depend on handcrafted rules and fixed assumptions, limiting their adaptability to evolving event dynamics and their ability to model intervention-aware opinion dynamics.

\subsection*{LLM-Based Agent Social Simulation}
Recent advances in LLMs~\cite{gpt-4-2023,Deepseek-r12025deepseek} have enabled cognitively enriched social simulations in which agents communicate, reason, and adapt through natural language. 
Representative systems such as \textit{Generative Agents}~\cite{Stanford-town-2023,Stanford1000agents-2024} demonstrate how LLM-driven agents~\cite{role2026-ACL2026} can support believable social reasoning and interaction in small-scale communities, while \textbf{S\textsuperscript{3}}~\cite{s3-2023}, \textit{SocioVerse}~\cite{Socioverse2025socioverse}, \textit{AgentSociety}~\cite{Agentsociety-2025agentsociety}, \textit{Yulan-OneSim}~\cite{Yulan-onesim2025-chenxu}, and \textit{OASIS}~\cite{Oasis-2025-shaojing} extend this line toward larger and more diverse social environments. 
Related evidence also suggests that LLMs can reproduce important structural properties of social systems, although they may still over-simplify some forms of heterogeneity~\cite{LLMs-generate-social-networks2024llms}. 
Despite this progress, most existing frameworks still under-model evolving intervention processes and the feedback loop between source-side actions and collective responses, which limits their ability to capture intervention-sensitive event evolution, opinion dynamics, and attitude shifts in real-world social networks.

\subsection*{Population-Aware Social Simulation}
Beyond micro-level interaction modeling, an important line of research studies how to represent public response and social dynamics at macro or population scales. 
In communication research, agenda-setting theory and related studies explain how issue salience, interpretive framing, and media influence shape collective attention and public cognition over time~\cite{Theory-Agenda-Setting1972agenda,agenda-setting1993evolution,network-agenda-setting2013toward,agenda-melding1999individuals}. 
Recent computational work further operationalizes these ideas in large-scale discourse analysis~\cite{mind-agenda-2015frame,Agenda-setting-Russiannews-2018-framing,Harmful-Agendas2023-towards,intermedia-agenda-setting-2023-rains}. 
Meanwhile, population-aware modeling has begun to move from individual-level simulation toward crowd-level public response prediction. 
For example, \textbf{GA-S\textsuperscript{3}}~\cite{GAS32025-ga} introduces group agents to improve scalability and population-level coherence in social network simulation, while \textit{MF-LLM}~\cite{MF-LLM2025mf} and \textit{PAC-LoRA}~\cite{PAC-LoRA-2025-ACMMM-chenxiuying} explore population-level response modeling. 
These studies highlight the value of macro-level representations for capturing large-scale collective trends, but still provide limited support for jointly modeling source-side intervention, public interaction, and intervention-sensitive opinion dynamics in a unified closed-loop framework.

\section{Methodology}
We formalize population-level decision dynamics (\S\ref{sec3.1:problem}) and introduce \textbf{IntervenSim}, an intervention-aware social simulation framework centered on intervention-aware simulation (\S\ref{sec3.2:Intervention-Aware Simulation}) and supported by representations for intervention dynamics (\S\ref{sec3.3:Supporting Representations}).


\subsection{Problem Definition}
\label{sec3.1:problem}
We aim to simulate large-scale social networks in text-based environments, where event dynamics arise from temporal interactions between source agents and crowd agents. In this work, \textit{event dynamics} refers to the temporal evolution of an event as manifested in the social network through intervention-sensitive engagement trajectories and group-level opinion dynamics, rather than the full real-world state of the event itself. The environment comprises $k$ source agents and a population of $N$ crowd agents $G=\{g_1,\dots,g_N\}$. At each timestep $t$, source agents produce an intervention vector $\vec{I}^{\,t}=(I_1^{\,t},\ldots,I_k^{\,t}) \in \mathcal{I}$ (\textbf{C1}), which influences the states and decisions of crowds. Each group agent represents a social cluster characterized by domain, country, interest relevance, and population size (\textbf{C2.1}). We denote the group states as $\vec{s}^{\,t}=(s_1^{\,t},\ldots,s_N^{\,t}) \in \mathcal{S}$ and their actions as $\vec{a}^{\,t}=(a_1^{\,t},\ldots,a_N^{\,t}) \in \mathcal{A}$, which jointly drive the temporal evolution of public response.

\begin{table}[h!]\small
\centering
\caption{Types of Source-side Interventions}
\begin{tabularx}{\linewidth}{
  @{\hspace{1.5pt}}p{0.5cm}
  @{\hspace{2pt}}p{1.7cm}
  @{\hspace{4.5pt}}p{5.2cm}
}
\toprule
\textbf{ID} & \textbf{Name} & \textbf{Description} \\
\midrule
\( I_1 \) & Prohibition & Restricts the spread of the event to reduce its visibility. \\
\( I_2 \) & Refutation & Counters misinformation to correct interpretive distortion. \\
\( I_3 \) & Publicity & Enhances event exposure to amplify public attention. \\
\( I_4 \) & Response & Provides institutional or opinion-leader feedback to shape interpretation. \\
\( I_5 \) & Announcement & Issues formal declarations or clarifications that reframe event meaning. \\
\bottomrule
\end{tabularx}
\label{tab1:interventions}
\end{table}

At each timestep $t$, the simulator outputs an engagement vector
\[
\mathbf{y}^{\,t}=\big(y_x^{\,t}\big)_{x \in \{\mathrm{view},\,\mathrm{like},\,\mathrm{comment},\,\mathrm{share}\}} \in \mathbb{R}^4,
\]
representing the predicted platform-level responses for each engagement type $x$, while the evolving group states encode the corresponding opinion dynamics and attitude shifts under intervention (\textbf{C2.2}). The objective is to generate an intervention-sensitive trajectory over $T$ timesteps that reproduces the real-world evolution of social response under event $\mathbf{e}$. In practice, we evaluate this objective primarily through the predicted view trajectory $\{y_{\mathrm{view}}^{\,t}\}_{t=1}^{T}$, while additionally analyzing whether the simulated process captures realistic opinion dynamics, attitude shifts, and consensus-polarization transitions. Compared with prior settings that assume a \textit{fixed event trajectory}, where the event context remains static once initialized, our setting treats event evolution as a closed-loop process in which source-side interventions, crowd deliberation, and collective feedback continuously reshape the social trajectory of the event over time.

\subsection{Intervention-Aware Simulation}
\label{sec3.2:Intervention-Aware Simulation}
To more realistically simulate social network dynamics, inspired by communication theories on attention guidance and interpretive framing, as well as classical theories of opinion dynamics~\cite{Theory-Agenda-Setting1972agenda,agenda-melding1999individuals,Theory-degroot-1974reaching,Theory-Hegselmann-Krause-2015opinion}, we design an \textbf{intervention-aware simulation} with two interactive processes:
(1) \textit{Source-Driven Process}, where media, government, institutions, and opinion leaders intervene in events to guide public attention and shape interpretation;
(2) \textit{Crowd Deliberation Process}, where crowd agents respond through interaction, reflection, and action, jointly driving social response and opinion dynamics over time.

\subsubsection*{Source-Driven Process.}  
Within the \textbf{IntervenSim} framework, source agents represent institutional or influential actors, such as media, government, public institutions, and opinion leaders, that affect event evolution through explicit intervention actions in the social network. Rather than assuming a fixed intended outcome, we model these actors as feedback-conditioned participants whose interventions depend on the current event context, the state of public response, and the evolving level of collective attention. Their interventions serve as observable public signals that can amplify visibility, provide clarification, offer interpretive guidance, issue formal responses, or restrict further diffusion.

\begin{figure}[ht]
    \centering
    \includegraphics[width=0.95\linewidth, height=0.75\linewidth]{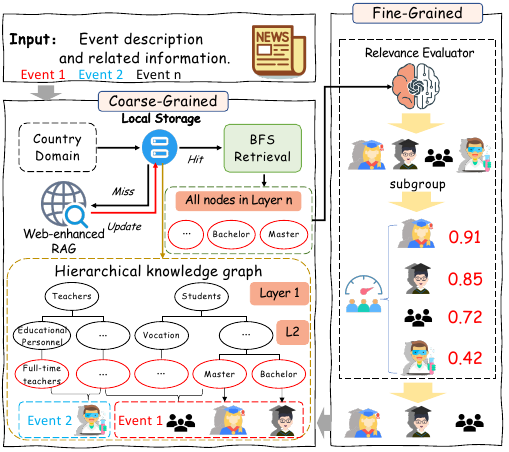}
    \caption{Dynamic group generation process. The system first performs coarse-grained grouping (country and domain) and then applies fine-grained specialization to generate event-specific subgroups. Top-$k$ $\cup$ ($Rel \ge 0.7$) subgroups retained for agent instantiation.}
    \label{fig3:dynamic-generation}
\end{figure}

At each timestep \( t \), each source agent selects an intervention \( I_k^t \in \mathcal{I} \) or remains inactive \( I_k^t = \emptyset \); the collection of these choices constitutes the intervention vector \(\vec{I}^{\,t} = (I_1^t, \ldots, I_k^t)\), which may itself be empty if no source-side action is taken (\textbf{C1}). Each intervention (Table~\ref{tab1:interventions}) modulates the ongoing social trajectory of the event by altering visibility, interpretive framing, or the subsequent direction of public response. Through repeated interaction with crowd-side deliberation, these source-side signals become part of a closed-loop process that continuously reshapes event and opinion dynamics over time. These interventions collectively shape two key dimensions of source-side influence:
\begin{itemize}[leftmargin=*]
\item \textbf{Attention guidance:} Source agents amplify selected events through publicity interventions, shaping \textbf{what the crowd thinks about}.  
\item \textbf{Interpretive framing:} Source agents modulate interpretation through interventions, shaping \textbf{how the crowd thinks about} them.   
\end{itemize}

\subsubsection*{Crowd Deliberation Process.}  
Crowd agents represent collective audiences that interpret and respond to authoritative signals through discussion and reflection.  
They engage in iterative exchanges of information and perspectives, enabling the bottom-up evolution of collective interpretation and public response through group-level reasoning.  
This process establishes the cognitive bridge between source-sid actions and public understanding within the evolving social network.

To model human-like decision-making and interaction among crowd agents, we design a dual-level reasoning mechanism that combines individual cognition with collective communication (\textbf{C2.2}). 
Each group agent $g_i$ maintains an internal \textit{cognitive state} $s_i^t$ and an evolving \textit{memory trace} $m_i^t$, both represented in natural language form.  
At each timestep $t$, the LLM-based Chain-of-Thought (CoT) policy $\pi_{\text{CoT}}$ generates the next action conditioned on the agent’s previous state, memory, the event context $\mathbf{e}^{t-1}$, the aggregated population signal $\bar{s}^{\,t-1}$, and source-side interventions $\vec{I}^{\,t}$:
\begin{equation}
a_i^t \sim \pi_{\text{CoT}}\big(s_i^{t-1},\, m_i^{t-1},\, \mathbf{e}^{t-1},\, \bar{s}^{\,t-1},\, \vec{I}^{\,t}\big)
\label{eq:action}
\end{equation}

After action generation, agents engage in peer communication based on the produced actions, then update their internal states and memories via an environment transition:
\begin{equation}
(s_i^t,\, m_i^t) \sim P\big(\cdot \mid s_i^{t-1},\, m_i^{t-1},\, a_i^t,\, \mathcal{C}_i^t\big)
\end{equation}
where $\mathcal{C}_i^t$ denotes the information exchanged with peers (e.g., shared attitudes, rationales).

Subsequently, the population signal is refreshed by aggregating updated states and actions:
\begin{equation}
\bar{s}^{\,t} \!\leftarrow\! \Phi\big(\{s_i^t,\, a_i^t\}_{i=1}^N\big)
\end{equation}
where $\Phi(\cdot)$ summarizes collective feedback into a global representation. This ordering captures how group-level updates precede population-wide aggregation, enabling a scalable and cognitively interpretable deliberation process.
Together, these processes form a unified closed-loop mechanism in which source-side interventions shape public attention and interpretation, and public responses in turn influence subsequent intervention and opinion dynamics.


\subsection{Supporting Representations}
\label{sec3.3:Supporting Representations}
To support intervention-aware simulation of event dynamics, 
we introduce two supporting representations for different stages of the simulation.
(1) \textit{Adaptive Crowd Representation}, which captures system states prior to interventions (pre);
(2) \textit{Dynamics Modeling with Intervention Feedback}, which models the temporal evolution of system states after interventions with explicit intervention feedback (post).

\subsubsection*{Adaptive Crowd Representation}
To enable adaptive and scalable simulation of public behavior, \textbf{IntervenSim} generates \textit{crowd agents} via a two-stage generation process (\textbf{C2.1}) that reflects how audiences self-organize around both general social backgrounds and event-specific relevance, as illustrated in Figure~\ref{fig3:dynamic-generation}. We adopt a top-down hierarchical generation strategy, combining coarse-grained semantic clustering with fine-grained population specialization, guided by Retrieval-Augmented Generation (RAG)~\cite{RAG-2020retrieval} and a relevance evaluation mechanism.



\noindent\textbf{Coarse-Grained Generation.} 
When a new event $\mathbf{e}$ occurs, we first extract its high-level semantic attributes, including domain $\mathbf{d}$ and country $\mathbf{c}$, via an LLM parser. 
Using $(\mathbf{d}, \mathbf{c})$ as retrieval keys, we obtain mid-level population templates from the hierarchical graph $\mathcal{G}$ through Breadth-First Search (BFS). 
If the corresponding branch is missing, web-enhanced RAG is used to retrieve public demographic and behavioral profiles relevant to the event context. 
These templates serve as coarse-grained population placeholders, providing stable structural priors for subsequent fine-grained specialization.


\definecolor{customgreen}{rgb}{0.239, 0.525, 0.288} 
\definecolor{darkblue}{rgb}{0.1, 0.2, 0.6}

\begin{algorithm}[t]
\caption{\textbf{IntervenSim Simulation Procedure}}
\label{alg:IntervenSim}
\small
\begin{algorithmic}[1]
\Require Event $\mathbf{e}$, $k$ source agents, $N$ crowd agents

\vspace{3pt}
\State \textbf{Initialize:} $\{s_i^0, m_i^0, A_i^0\}_{i=1}^{N}$; set $\bar{s}^{\,0} \!\leftarrow\! \Phi(\{s_i^0, A_i^0\})$

\vspace{3pt}
\For{$t = 1$ to $T$}
    \State Source agents select interventions: 
    
    $\vec{I}^{\,t}\!\leftarrow\! (I_1^t, \dots, I_k^t)$, where $I_k^t \!\in\! \mathcal{I}$ or $I_k^t{=}\emptyset$

    \vspace{3pt}
    \State Each crowd agents samples action:
    
    $a_i^t \!\sim\! \pi_{\text{CoT}}(s_i^{t-1}, m_i^{t-1}, \mathbf{e}^{t-1}, \bar{s}^{\,t-1}, \vec{I}^{\,t})$

    \vspace{3pt}
    \State Public agents communicate with peers $\mathcal{N}(i,t)$ to form $\vec{D}^{\,t}$ and $A_i^t$
    \small{
    \[
    A_i^t =
    \begin{cases}
    \pi_{\text{CoT}}\big(A_i^{t-1},\, m_i^{t-1},\, \vec{D}^{\,t},\, \vec{I}^{\,t}\big), 
    & \text{if } \vec{D}^{\,t} \text{ or } \vec{I}^{\,t} \neq \emptyset, \\[6pt]
    \frac{1}{|\mathcal{N}_\epsilon(i,t)|} \sum_{j \in \mathcal{N}_\epsilon(i,t)} A_j^{t-1},
    & \text{otherwise.}
    \end{cases}
    \]
    }
    \State Define communication signal $\mathcal{C}_i^t \!\leftarrow\! \{A_i^t,\, \vec{D}^{\,t}\}$

    \vspace{2pt}
    \State Update internal states:
    
    $(s_i^t, m_i^t) \!\sim\! P(\cdot \mid s_i^{t-1}, m_i^{t-1}, a_i^t, \mathcal{C}_i^t)$

    \vspace{3pt}
    \State Aggregate $\bar{s}^{\,t} \!\leftarrow\! \Phi(\{s_i^t, a_i^t\}_{i=1}^{N})$

    \vspace{3pt}
    \State Compute engagement vector based on $\bar{s}^{\,t}$:
    
    $\mathbf{y}^{\,t} = \big(y_x^{\,t}\big)_{x \in \{\mathrm{view},\, \mathrm{like},\, \mathrm{comment},\, \mathrm{share}\}} \in \mathbb{R}^4$

    \vspace{3pt}
    \State Update event-level trajectory $\mathbf{e}^{\,t} \!\leftarrow\! \mathbf{e}^{\,t-1} \cup \{\mathbf{y}^{\,t}\}$
\EndFor

\vspace{3pt}
\State \textbf{Output:} Trajectory $\{\mathbf{y}_{\mathrm{view}}^{\,t}\}_{t=1}^{T}$ representing the simulated temporal evolution of engagement under event $\mathbf{e}$
\end{algorithmic}
\end{algorithm}

\noindent\textbf{Fine-Grained Specialization.} 
While coarse categories provide structural stability, real-world events require finer resolution. 
We introduce a fine-grained specialization step using a \textbf{Relevance Evaluator}, which conditions on the retrieved coarse-grained population templates and the event $\mathbf{e}$ to generate a candidate subgroup pool $\mathcal{C}_i = \{c_1, \dots, c_J\}$ for each coarse-grained population node. 

Each candidate subgroup $c_j \in \mathcal{C}_i$ receives a relevance score:
{{\small
\[
s_j = \mathrm{Rel}(c_j, \mathbf{e}) 
= \lambda \frac{\langle \phi(c_j), \phi(\mathbf{e}) \rangle}{\|\phi(c_j)\|\|\phi(\mathbf{e})\|} 
+ (1-\lambda)\psi(c_j, \mathbf{e})
\]
}}
where $\phi(\cdot)$ is an embedding-based semantic encoder and $\psi(\cdot,\cdot)$ is an LLM prompt-based relevance score normalized to $[0,1]$, with $\lambda \in [0,1]$ balancing the two terms.  

The selected fine-grained subgroup set is defined as
\begin{equation*}
\mathcal{G}_i^{\text{fine}} = \mathrm{TopK}\big(\{c_j\}_{j=1}^{J}, k\big) 
\cup \big\{c_j \mid s_j \ge 0.7 \big\},
\end{equation*}
ensuring at least $\max\{k,\,|\{j:s_j\ge0.7\}|\}$ specialized subgroups for each coarse-grained population node.  
The selected subgroup sets are then instantiated as fine-grained crowd agents, which collectively form the updated $G = \{g_1, \dots, g_N\}$ used for subsequent simulation, enabling adaptive population refinement according to event-specific salience and heterogeneous discourse contexts.

\begin{table*}[h!]
\centering
\caption{Comparison results of all methods. The second-best score is underlined, and the best one is bold.}
\setlength{\tabcolsep}{6.5pt}
\begin{tabular}{
>{\raggedright\arraybackslash}p{3.1cm}
>{\centering\arraybackslash}p{1.4cm}
>{\centering\arraybackslash}p{1.7cm}
>{\centering\arraybackslash}p{1.7cm}
>{\centering\arraybackslash}p{1.4cm}
>{\centering\arraybackslash}p{1.4cm}
>{\centering\arraybackslash}p{2.0cm}
}
\toprule
\multirow{2}*{\textbf{Method}} & \multirow{2}*{\textbf{$W_1\downarrow$}} & \multirow{2}*{\textbf{MAPE~$\downarrow$}} & \multicolumn{2}{c}{\textbf{DTW}} & \multirow{2}*{\textbf{Z-score}} & \multirow{2}*{\textbf{Ave Agent Num}} \\
\cmidrule(lr){4-5}
& & & Mean(e+07)~$\downarrow$ & Std~$\downarrow$ & & \\
\midrule
PSP~\citep{PSP-2018}          & 0.2522 & 69.12\% & 3.40 & 0.4207 & --    & --    \\
S$^3$~\citep{s3-2023}         & 0.2519 & 68.66\% & 3.09 & 0.4035 & 1.22  & 1000  \\
GA-S$^3$~\citep{GAS32025-ga}  & \underline{0.1477} & \underline{16.48\%} & \underline{1.30} & \underline{0.1890} & 0.81  & 16    \\
\rowcolor{gray!20}
\textbf{IntervenSim (Ours)}   & \textbf{0.1159} & \textbf{9.63\%} & \textbf{0.43} & \textbf{0.1105} & \textbf{0.73} & \textbf{10} \\
\scriptsize & \scriptsize($\downarrow$21.53\%) & \scriptsize($\downarrow$41.6\%) & \scriptsize($\downarrow$66.9\%) & \scriptsize($\downarrow$41.6\%) & & \\
\bottomrule
\end{tabular}
\label{tab2:compare1}
\end{table*}

\subsubsection*{Opinion Dynamics Modeling}

To model realistic opinion evolution, we integrate discussion-based interactions with hybrid updating dynamics inspired by \textit{Opinion Dynamic Model}~\cite{Theory-degroot-1974reaching, Theory-Hegselmann-Krause-2015opinion}. 
At each timestep, each group agent holds an attitude state $A_i^t$ representing its evaluative stance toward the event.  
Agents interact with peers selected according to their interaction parameter $\mathcal{N}_\epsilon(i,t)$ (determined by prior actions $a_i^t$) through comment-based discussions that transmit peer opinions and contextual content. 
When a discussion signal $\vec{D}^{\,t}$ or intervention signal $\vec{I}^{\,t}$ is detected, attitude updating follows a cognitive reasoning path; otherwise, it evolves through a dynamical model of social influence:

{\small
\[
A_i^t =
\begin{cases}
\pi_{\text{CoT}}\big(A_i^{t-1},\, m_i^{t-1},\, \vec{D}^{\,t},\, \vec{I}^{\,t}\big), 
& \text{if } \vec{D}^{\,t} \text{ or } \vec{I}^{\,t} \neq \emptyset, \\[6pt]
\frac{1}{|\mathcal{N}_\epsilon(i,t)|} \sum_{j \in \mathcal{N}_\epsilon(i,t)} A_j^{t-1},
& \text{otherwise.}
\end{cases}
\]
}
where $\epsilon \in [0,1]$, and the detailed formulation of the opinion dynamics is provided in Appendix~\ref{appB:opinion_dynamics}.
The corresponding communication signal is defined as: $\mathcal{C}_i^t \leftarrow\ \{A_i^t,\, \vec{D}^{\,t}\}$ (\textbf{C2.2}).
This formulation couples reflective reasoning and continuous social influence within the same transition function $P$, ensuring cognitive interpretability while preserving realistic opinion dynamics. Pseudocode~\ref{alg:IntervenSim} summarizes the \textbf{IntervenSim} framework.

\section{Experiment}
\subsection{Settings}

\noindent\textbf{Model.}
We adopt the locally deployed \texttt{LLaMA3-8B}~\cite{llama3-2024} as our base large language model, and additionally use \texttt{Qwen2.5-7B}~\cite{Qwen2.5-2025qwen2} for ablation analysis. To ensure deterministic and stable outputs during all reasoning and decision-making stages, we fix the decoding temperature at~0.1. For hierarchical generation of group agents conditioned on event contexts, we employ the remote \texttt{GPT-4o}~\cite{gpt-4-2023} API, which dynamically constructs agent profiles through prompt-based few-shot generation.

\noindent\textbf{Dataset.}
We adopt the \textbf{S}ocial \textbf{N}etwork \textbf{B}enchmark (SNB), originally introduced by GA-S$^3$~\cite{GAS32025-ga}, to ensure comparability with prior simulations. SNB comprises real-world events from Twitter (X), Reddit, and Weibo, each annotated with title, content, metadata, and 7-day network traffic. The dataset covers diverse domains and regions, including both sudden incidents and long-term events, and contains cases involving consensus and polarization phenomena. Expert-curated annotations on attitude shifts and behavioral trends provide external references for evaluating traffic fit, opinion evolution, and group-level behavioral diversity.

\begin{figure}[ht]
    \centering
    \includegraphics[width=\linewidth, height=0.68\linewidth]{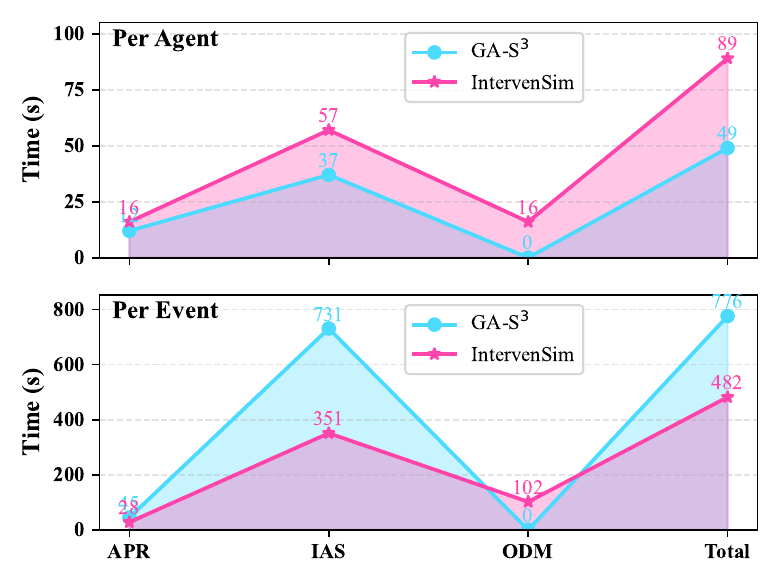}
    \caption{Time cost comparison between GA-S$^3$ and IntervenSim across different stages. Circles represent GA-S$^3$, stars represent IntervenSim, solid lines indicate per-agent time, and dashed lines indicate per-event time.}
    \label{fig4:timecost}
\end{figure}

\noindent\textbf{Evaluation Metrics.}
We adopt a unified evaluation protocol and report results using the following metrics:
\begin{itemize}[leftmargin=*]
\item \textbf{Wasserstein Distance ($W_1$)}~\cite{w1-2009wasserstein}: 
Evaluates the \textit{distributional difference} between simulated and real engagement behavior. 
Lower $W_1$ indicates closer alignment. 
Empirically, $W_1 < 0.15$ denotes high fidelity, $0.15\!-\!0.35$ indicates moderate deviation, and $W_1 > 0.35$ reflects a clear mismatch in behavioral distribution.
\item \textbf{Mean Absolute Percentage Error (MAPE)}~\cite{mape-1999}: 
Measures the relative prediction error of engagement magnitudes, evaluating the accuracy of the four-dimensional engagement vector \( \mathbf{y}^{\,t} \in \mathbb{R}^4 \).
\item \textbf{Dynamic Time Warping (DTW)}~\cite{DTW-2007}: 
Assesses temporal alignment between simulated and real \textit{view traffic trajectories} \( \{y_{\mathrm{view}}^t\} \), evaluating the model’s ability to capture realistic fluctuation and trend dynamics.
\item \textbf{Z-score (Reproducibility)}~\cite{z-score-2016}: 
Reports the stability of results across repeated simulations. 
We run each configuration \textbf{three times}, and report the standardized deviation; $|Z| < 1$ indicates high reproducibility.
\end{itemize}

\noindent\textbf{Baselines.}
We compare our method against the following representative approaches:
\begin{itemize}[leftmargin=*]
    \item \textbf{PSP}~\cite{PSP-2018}: Identifies popularity stage patterns in social media and applies pattern-matching techniques to forecast future trends.
    
    \item \textbf{S\textsuperscript{3}}~\cite{s3-2023}: Uses LLM-based individual agents to simulate virtual social networks with enhanced perception and reasoning.

    \item \textbf{GA-S$^3$}~\cite{GAS32025-ga}: A social simulation framework using static group agents with sentiment and memory dynamics to model network traffic and forecast large-scale events.
\end{itemize}

\begin{table*}[h!]\small
\centering
\caption{Ablation experiments. \(\ell\)1 corresponds to the results of GA-S$^3$, \(\ell\)8* uses Qwen2.5-7b~\cite{Qwen2.5-2025qwen2}. Default settings are marked in gray.}
\label{tab2:ablation}
\setlength{\tabcolsep}{5.4pt}
\renewcommand{\arraystretch}{1.08}
\begin{tabular}{
>{\centering\arraybackslash}m{1.9cm}
>{\centering\arraybackslash}m{0.72cm}
>{\centering\arraybackslash}m{0.72cm}
>{\centering\arraybackslash}m{0.72cm}|
>{\centering\arraybackslash}m{1.32cm}
>{\centering\arraybackslash}m{1.32cm}
>{\centering\arraybackslash}m{1.32cm}
>{\centering\arraybackslash}m{1.32cm}
>{\centering\arraybackslash}m{1.32cm}
>{\centering\arraybackslash}m{1.52cm}
}
\toprule
\multirow{2}*{\(\ell\)} & \multirow{2}*{APR} & \multirow{2}*{IAS} & \multirow{2}*{ODM} & \multirow{2}*{$W_1~\downarrow$} & \multirow{2}*{MAPE~$\downarrow$} & \multicolumn{2}{c}{DTW} & \multirow{2}*{Z-score} & \multirow{2}*{Ave Agent Num} \\
\cmidrule(lr){7-8}
&&&&&& Mean~$\downarrow$ & Std~$\downarrow$ & \\
\midrule
1 (GA-S$^3$) & --         & --          & --          & 0.1477 & 16.48\% & 1.30e+07 & 0.1890 & 0.81 & 16 \\
2            & \checkmark & --          & --          & 0.1993 & 16.90\% & 1.33e+07 & 0.1903 & 0.74 & 6 \\
3            & --         & \checkmark & --          & 0.1389 & 15.59\% & 9.01e+06 & 0.1862 & 0.83 & 20 \\
4            & --         & --          & \checkmark & 0.1277 & 16.12\% & 0.79e+07 & 0.1873 & 0.89 & 16\\
5            & \checkmark & \checkmark & --          & 0.1526 & 12.03\% & 5.05e+06 & 0.1808 & 0.76 & 10 \\
6            & \checkmark & --          & \checkmark & 0.1444 & 14.72\% & 8.30e+06 & 0.1854 & 0.78 & 6 \\
7            & --         & \checkmark & \checkmark & \textbf{0.1111} & 13.48\% & 4.91e+06 & 0.1578 & 0.84 & 20\\
\midrule
8 (qwen2.5-7b) & \checkmark & \checkmark & \checkmark & 0.1171 & 10.16\% & 4.47e+06 & 0.1217 & 0.76 & 10\\
\rowcolor{gray!30}
9 (IntervenSim) & \checkmark & \checkmark & \checkmark & \underline{0.1159} & \textbf{9.63\%} & \textbf{4.30e+06} & \textbf{0.1105} & \textbf{0.73} & 10\\
\bottomrule
\end{tabular}
\end{table*}

\subsection{Comparison with Baselines}

The main results are summarized in Table~\ref{tab2:compare1}, from which several key observations can be made.

\textbf{IntervenSim consistently outperforms all baselines across all evaluation metrics.} 
Compared to the previous state-of-the-art, IntervenSim shows a 21.53\% improvement in $W_1$ and a 41.6\% improvement in MAPE, indicating more accurate alignment with real engagement distributions. 
DTW mean is reduced by 66.9\% (with a 41.6\% reduction in variance), showing substantially improved tracking of temporal traffic dynamics.
Z-scores remain within $\pm 1$ across three repeated runs, confirming stable and reproducible simulation behavior.
Overall, these results suggest that IntervenSim more faithfully captures the evolution of public attention and interaction patterns in social networks.

\textbf{IntervenSim shows marked advantages in modeling realistic network traffic variations.} While prior baselines can reflect overall trends, they often struggle with fine-grained temporal shifts and multi-phase transitions typical in real-world social events. IntervenSim reduces the DTW mean by an order of magnitude (from $10^7$ to $10^6$), indicating substantially improved alignment with temporal patterns. Importantly, it achieves these gains with fewer agents on average (10 in ours vs. 16 in GA-$S^3$), highlighting the framework’s computational efficiency and its ability to accurately track complex social dynamics at scale.


\subsection{Ablation Study}
We perform ablation experiments to evaluate the role and efficiency of the three core modules:

\begin{figure}
    \centering
    \includegraphics[width=\linewidth, height=0.68\linewidth]{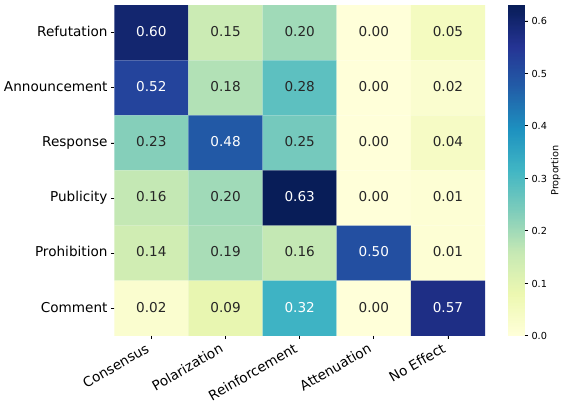}
    \caption{Heatmap of five intervention and comment types versus opinion transitions.}
    \label{fig7:heatmap}
\end{figure}

\begin{figure*}
    \includegraphics[width=\linewidth]{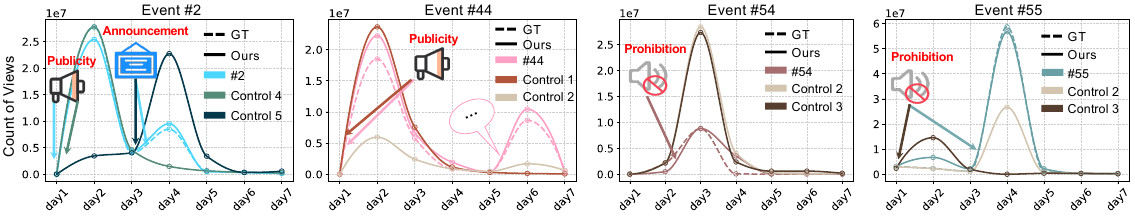}
    \caption{Traffic trend comparison under different ablation settings across events. Results show that interventions, their timing, and the presence of comments significantly influence traffic dynamics.}
    \label{fig5:intervene_control}
\end{figure*}

\begin{figure*}
    \includegraphics[width=\linewidth]{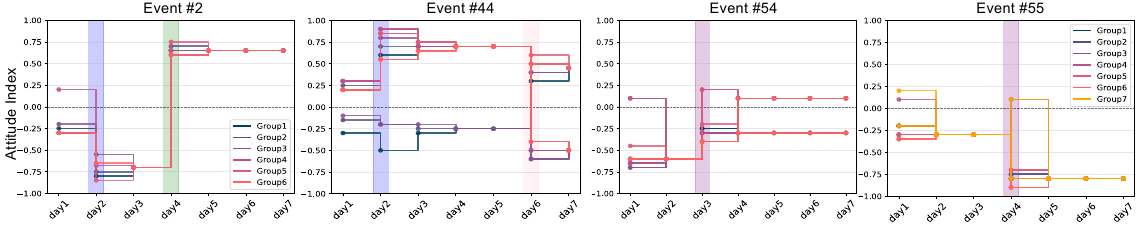}
    \caption{Opinion evolution of different group agents across four events. Consensus, polarization, attenuation, and reinforcement patterns emerge. Opinion shifts commonly follow interventions or high-impact public comments and align with surges in event traffic. Shaded areas indicate periods of intervention or high-impact comment activity.}
    \label{fig6:attitude}
\end{figure*}

\begin{enumerate}[leftmargin=*]
  \item \textbf{Module Effectiveness.}
  We individually and jointly disable each module to evaluate its contribution to overall performance.
  
  \item \textbf{Computational Efficiency.}
  We measure the runtime overhead introduced by each module to analyze its impact on overall efficiency.
\end{enumerate}

\noindent\textbf{Q1: Impact of Core Modules.}
To answer \textbf{Q1}, we perform ablations over the three core modules:

\begin{itemize}[leftmargin=*]
  \item \textbf{IAS (Intervention-Aware Simulation):} We remove source-driven intervention and let crowd agents evolve solely through peer deliberation.
  \item \textbf{APR (Adaptive Public Representation):} We replace event-adaptive group construction with a fixed predefined group structure.
  \item \textbf{ODM (Opinion Dynamics Modeling):} We disable agent discussion and attitude evolution, preventing cross-agent influence.
\end{itemize}

The ablation results are summarized in Table~\ref{tab2:ablation}. 
Comparing \(\ell\)4 to \(\ell\)1 shows that enabling \textbf{ODM} significantly improves temporal alignment, indicating the importance of multi-agent attitude evolution. 
\textbf{APR} benefits performance primarily when combined with other modules (\(\ell\)5, \(\ell\)6, \(\ell\)9), as event-adaptive group structures help reduce mismatch between agent roles and event semantics; however, using APR alone (\(\ell\)2) can slightly degrade performance due to reduced agent diversity. 
Removing the \textbf{IAS} module (\(\ell\)3, \(\ell\)6) leads to broad declines across all metrics, highlighting the necessity of source-driven agenda formation in stabilizing deliberation. 
The full model (\(\ell\)9), integrating all three modules, achieves the best overall results. 
Additionally, experiments using Qwen2.5-7B (\(\ell\)8) show consistent gains, demonstrating that the improvements originate from the framework itself rather than a specific LLM backbone.

\noindent\textbf{Q2: Impact of Time Cost and Module Efficiency.}  
IntervenSim introduces more complex computations per agent than GA-S$^3$, primarily due to its enhanced \textbf{source-driven influence} and \textbf{group-level interaction dynamics}, which reduce large-scale redundancy at the system level (figure~\ref{fig4:timecost}).

Each module contributes the following time costs: the \textbf{APR} module adds 4 seconds per agent, the \textbf{IAS} module adds 20 seconds, and the \textbf{ODM} module adds 16 seconds for inter-agent interactions. Despite the increased per-agent time (49 seconds $\rightarrow$ 89 seconds, an 81.63\% increase), \textbf{IntervenSim reduces total simulation time per event} (776 seconds $\rightarrow$ 482 seconds, a 37.89\% reduction) due to the smaller number of agents required.

\subsection{Case Study}
\subsubsection*{Intervention Analysis.}  
To evaluate the effects of the intervention-aware closed-loop mechanism, we conduct \textit{controlled experiments} on (i) intervention timing in the \textit{Source-Driven Process} and (ii) the presence of communicative responses in the \textit{Crowd Deliberation Process}. Each experiment compares the observed (real-timestamped) condition with event-matched counterfactual controls:

\begin{itemize}[leftmargin=*]
    \item \textbf{Control 1:} Disabling public commenting
    \item \textbf{Control 2:} Removing all source-side interventions
    \item \textbf{Control 3:} Changing the intervention time
    \item \textbf{Control 4:} Removing the second intervention
    \item \textbf{Control 5:} Removing the first intervention
\end{itemize}

\textbf{IntervenSim accurately models how intervention type and timing influence collective attention through source-side interventions and public response.}  
As shown in Figure~\ref{fig5:intervene_control}, IntervenSim successfully reproduces a wide range of intervention effects observed in real-world social dynamics across four representative events (detailed descriptions and characteristics are provided in the Appendix table~\ref{tab5:case event}).  
Interventions such as \textit{Publicity}, \textit{Announcement}, \textit{Response}, and \textit{Refutation} amplify engagement (e.g., Events~\#2 and~\#44); when occurring sequentially, they generate bi-modal or recursively rising attention patterns, as both source-side signaling and subsequent \textit{public discussion} can trigger renewed waves of participation, reflecting how exposure and discourse jointly sustain collective attention.  
In contrast, \textit{Prohibition} interventions (e.g., Events~\#54 and~\#55) produce either effective suppression or rebound effects, depending on contextual timing and communicative openness.  
More detailed event-level analyses are provided in the Appendix~\ref{appB.3:controlled intervention}.

\subsubsection*{Opinion Dynamics.}  
We categorize collective attitude evolution into four primary modes:  

\begin{itemize}[leftmargin=*]
    \item \textbf{Polarization:} agents divide into opposing clusters of positive and negative attitudes.
    \item \textbf{Consensus:} attitudes converge toward a dominant shared orientation.
    \item \textbf{Reinforcement:} existing attitudes intensify while retaining their original polarity.
    \item \textbf{Attenuation:} attitude magnitudes weaken, indicating reduced conviction.
\end{itemize}

\textbf{IntervenSim captures the co-evolution of external interventions and crowd deliberation, reproducing how collective attitudes shift and reorganize through interactive communication dynamics.}  
As shown in Figure~\ref{fig6:attitude}, across representative events, IntervenSim successfully reproduces these communicative transformations observed in real-world discourse.  
Sequential \textit{Publicity} and \textit{Announcement} interventions (e.g., Event~\#2) trigger alternating consensus shifts from negative to positive orientations, producing recursive engagement peaks.  
Single \textit{Publicity} actions followed by public commentary (e.g., Event~\#44) generate cycles of reinforcement and polarity reversal, showing how internal deliberation reshapes communicative balance.  
\textit{Prohibition} interventions (e.g., Events~\#54 and~\#55) yield contrasting outcomes: timely restriction attenuates negative reproduction, whereas delayed enforcement amplifies critical narratives due to entrenched communicative momentum.  
Notably, \textit{days with major attitude realignments coincide with traffic peaks, and peak amplitude correlates with the magnitude of collective attitude shifts, indicating a tight coupling between opinion evolution and public attention.}  
More detailed event-wise analyses are provided in the Appendix~\ref{appB.4:opinion dynamics}.

\subsubsection*{Intervention–Opinion Correlation.}  
As shown in Figure~\ref{fig7:heatmap}, \textit{consensus} commonly follows \textbf{announcements} or \textbf{refutations}, while \textbf{responses} often trigger \textit{polarization}. \textbf{Publicity} tends to \textit{reinforce} existing attitudes, and \textbf{prohibitions} usually \textit{attenuate} them, though excessive restriction can provoke backlash and renewed polarization. \textbf{Comments}, reflecting diverse perspectives, have limited direct impact but may reinforce echo-chamber effects.  
Overall, these results suggest that \textbf{different intervention types tend to guide collective attitudes toward distinct evolutionary patterns within public response}.  

\section{Conclusion}
IntervenSim presents a large-scale and scalable framework for simulating social network dynamics with interpretable structure and empirical realism. By integrating intervention-aware simulation, source-crowd interaction, and opinion dynamics modeling, it captures how interventions and crowd deliberation jointly shape collective attention and opinion evolution. Experimental results demonstrate that IntervenSim reproduces real-world patterns of amplification, suppression, consensus, and polarization with higher accuracy and efficiency than prior frameworks. This paradigm bridges LLM-based social agents with established social theories, providing a unified foundation for future research on interpretable, data-aligned, and dynamically adaptive social simulations.

\begin{acks}
To Robert, for the bagels and explaining CMYK and color spaces.
\end{acks}

\bibliographystyle{ACM-Reference-Format}
\bibliography{main}

\appendix

\clearpage
\appendix

\let\oldtwocolumn\twocolumn
\renewcommand\twocolumn[1][]{%
    \oldtwocolumn[{#1}{
    \begin{center}
    \vspace{-5.5em}
    \includegraphics[width=\textwidth]{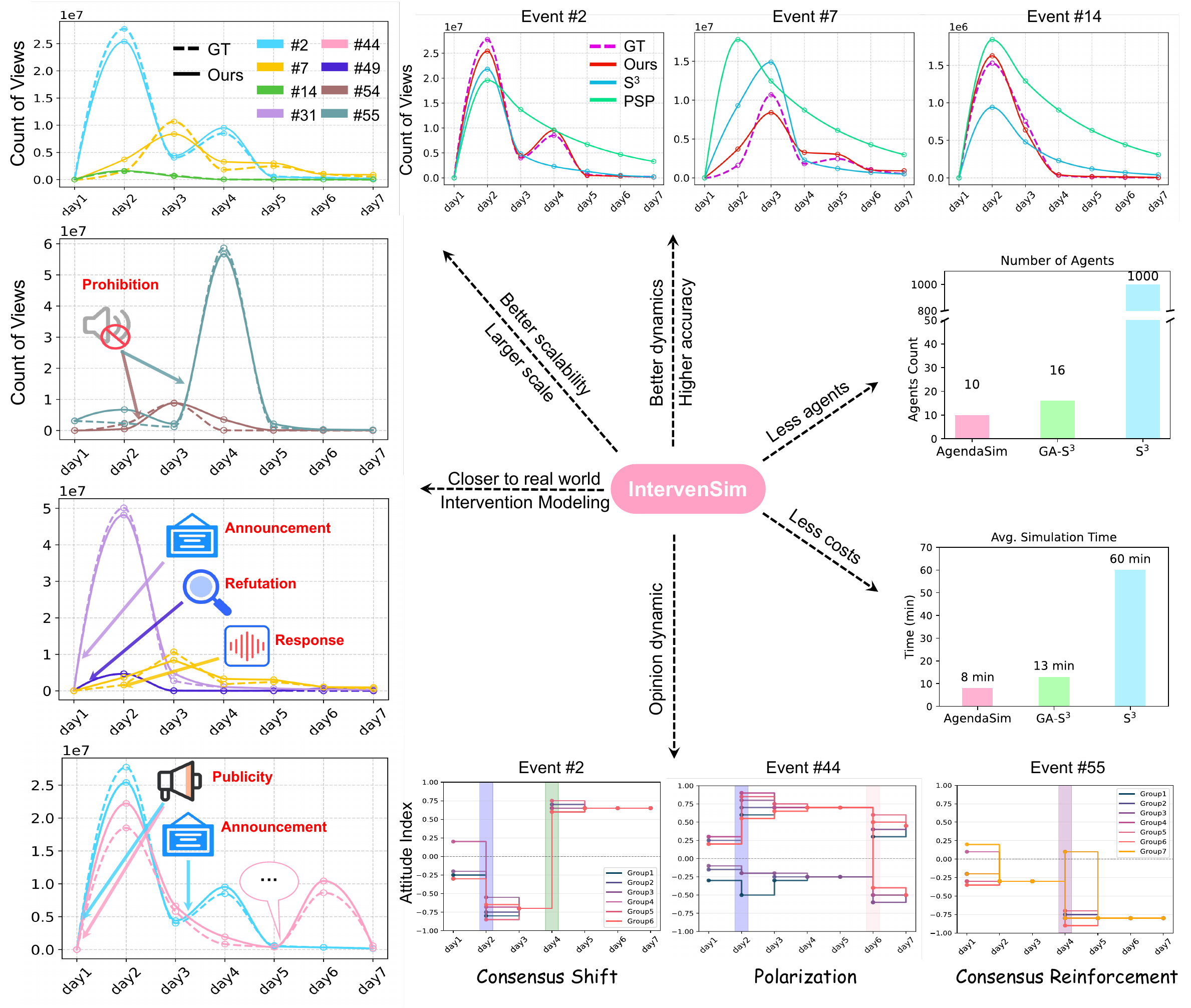}
    \captionof{figure}{
        Overall intuitive impact of our framework.
        The results demonstrate the advantages of \textbf{IntervenSim}, including 
        \textbf{high accuracy}, 
        \textbf{robust opinion dynamics and intervention modeling}, 
        \textbf{greater real-world fidelity}, 
        \textbf{large-scale generalizability and scalability}, 
        and \textbf{lower simulation time cost}.
    }
    \label{fig8:intuitive impact}
    \end{center}
    }]
}

\twocolumn[
{\Large\centerline{\textbf{Appendix}}} 
\vspace{6.5em}
]
As shown in Fig.~\ref{fig8:intuitive impact}, \textbf{IntervenSim} demonstrates superior performance across multiple aspects. This appendix provides supplementary discussions and implementation details that complement the main paper, including:

\section*{Appendix Overview}
\noindent\textbf{A.} \hyperref[appA:related_work]{Related Work}\par
\noindent\textbf{B.} \hyperref[appB:opinion_dynamics]{Opinion Dynamics Analysis}\par
\noindent\textbf{C.} \hyperref[appC:experimental_details]{Detailed Experimental Setup}\par
\noindent\textbf{D.} \hyperref[appD:ethical]{Ethical Considerations}\par
\noindent\textbf{E.} \hyperref[appE:prompt]{Prompt Design}\par

\section{Related Work}
\label{appA:related work}

\subsection*{Social Simulation Systems}

Traditional social simulation systems~\cite{SIR-model2002spread,NC-Social-Media-2022universality} model how individual behaviors collectively shape societal structures~\cite{Social-Simulation-overview-2014,social-network-analysis2004development,Mvp2025-acmmm}.  
Existing paradigms can be grouped into three main categories:  

\textit{(1) Mechanistic models.}  
These approaches describe collective behavior through explicit equations or procedural rules, including system-dynamics~\cite{traditional-simulation-system-dynamics,Systems-Based-Approach-2012} and discrete-event simulations~\cite{traditional-simulation-discrete-events,dynamics-lupeng-2021swarm}.  
They capture macro-level diffusion processes such as epidemic spreading~\cite{SIR-model2002spread} and online mobilization~\cite{5rule-lupeng-2018exploring,peak-time-lupeng2018big,ACMMM2025-Stealthy-AE}, but rely on fixed parameters and lack adaptive cognition.  

\textit{(2) Empirical and statistical models.}  
These data-driven frameworks identify regularities in social diffusion, such as PSP~\cite{PSP-2018} and peak-based participation dynamics~\cite{peak-height-lupeng-2018predicting,shapes-lupeng2019strength}, revealing consistent patterns like the “5‰ rule” of participation~\cite{5rule-lupeng-2018exploring}.  
While effective for large-scale pattern discovery, they offer limited explanatory power for interaction mechanisms underlying opinion dynamics.

\textit{(3) Agent-Based Social Simulation.}  
To overcome the limitations of aggregate and rule-based models, \textbf{agent-based social simulation (ABSS)} offers a bottom-up approach to studying complex social phenomena~\cite{Old-Agent-Based-Social-Simulation-2002,squazzoni2008micro,detection2025aaai}.  
By modeling heterogeneity, autonomy, and interaction among individuals, ABSS allows macro-level patterns to emerge from local decisions and cognitive mechanisms~\cite{Multiagent-Systems2005,Multi-Agent-Systems-application-2018}.  
Classical works such as Schelling’s segregation model~\cite{first-agent-based-model-schelling-1971,Cellular-Automata-1998,Micromotives-Macrobehavior-2006} and Epstein \& Axtell’s \textit{Sugarscape}~\cite{first-large-scale-agent-model-1996} illustrate how simple preferences can yield complex societal structures, while later studies incorporated organizational and normative constraints~\cite{Multiagent-Systems2005}.  
However, traditional ABMs rely on handcrafted rules and static environments, limiting their ability to model dynamic evolution and intervention effects, and motivating LLM-driven simulations for scalable, adaptive social dynamics.

\subsection*{LLM-Based Agent Social Simulation}
Recent advances in LLMs and multimodal AI~\cite{Coupled-mamba2024-nips,Qwen2.5-2025qwen2,LPT-2026logical,LoRA-Mixer-2025-ICLR,Sf2t2025-cvpr,tracking2025aaai} have transformed agent-based social simulation~\cite{ACMMM-2025-simulatior,ICML2025position}, enabling agents to communicate, reason, and adapt in natural language~\cite{LLMs-generate-social-networks2024llms,Stanford-town-2023,Stanford1000agents-2024}.  
Early works such as \textit{Generative Agents}~\cite{Stanford-town-2023,ACMMM2025-genagent} and S\textsuperscript{3}~\cite{s3-2023} demonstrated lifelike social reasoning and interaction~\cite{ACMMM2025-CoA,dai2025psycher1reliablepsychologicalllms,long2026emo}, establishing the feasibility of LLM-driven social simulation in small-scale or medium-scale environments.  
Subsequent frameworks such as \textit{SocioVerse}~\cite{Socioverse2025socioverse}, \textit{AgentSociety}~\cite{Agentsociety-2025agentsociety}, \textit{Yulan-OneSim}~\cite{Yulan-onesim2025-chenxu}, and \textit{OASIS}~\cite{Oasis-2025-shaojing} further pushed this line toward more scalable infrastructures, richer agent coordination, and larger social environments, including simulations with up to one million agents.  
Meanwhile, task-oriented systems such as \textit{PopSim}~\cite{PopSim-2025-liuwu} explore how LLM-based social network simulation can be applied to social media popularity prediction, extending agent simulation from open-ended social interaction to event-centered predictive settings.  
Collectively, these studies substantially improve realism, scalability, and semantic richness in social simulation. However, most existing frameworks still under-model evolving intervention processes and feedback loops between source-side actions and collective responses.  
As a result, they remain limited in capturing intervention-sensitive event evolution, opinion dynamics, and attitude shifts in real-world social networks, motivating the design of \textbf{IntervenSim}.

\subsection*{Population-Aware Social Simulation}

Beyond micro-level interaction modeling, an important line of research studies how to represent public response and social dynamics at macro or population scales.  
In communication research, agenda-setting theory~\cite{Theory-Agenda-Setting1972agenda,agenda-setting1993evolution,network-agenda-setting2013toward,agenda-melding1999individuals} and its extensions provide a conceptual account of how issue salience, interpretive framing, and media influence shape collective attention and public cognition over time~\cite{Public-opinion1922public,NAACL-2025-trendsim,wang2026mindreasoning,wang2026prismopinion}.  
Recent computational studies further operationalize these ideas in large-scale discourse analysis, showing how framing, salience, and intermedia influence can be modeled from text and temporal signals~\cite{mind-agenda-2015frame,Agenda-setting-Russiannews-2018-framing,Harmful-Agendas2023-towards,intermedia-agenda-setting-2023-rains}.  

At the same time, emerging work has begun to move from individual-level simulation toward group-, crowd-, or population-level response modeling~\cite{ACMMM2025-SegTraj,ACMMM2025-crowdSim}.  
For example, \textbf{GA-S\textsuperscript{3}}~\cite{GAS32025-ga} introduces \textit{Group Agents} to improve scalability and population-level coherence in social network simulation, while \textit{MF-LLM}~\cite{MF-LLM2025mf} incorporates mean-field approximation into LLM-based simulation to improve scalability at the population level.  
Related work such as \textit{PAC-LoRA}~\cite{PAC-LoRA-2025-ACMMM-chenxiuying} explicitly models public response across individual and crowd levels, further highlighting the value of macro-level representations for capturing large-scale collective trends.  
However, these studies still provide limited support for jointly modeling source-side intervention, public interaction, and intervention-sensitive opinion dynamics within a unified closed-loop simulation framework.  
Our work builds on this broader macro-level perspective while focusing on how source-side interventions and collective responses co-evolve in social network events.

\definecolor{color1}{HTML}{C0C0EC}  
\definecolor{color2}{HTML}{D1E5CE}  
\definecolor{color3}{HTML}{FDF2F5}  
\definecolor{color4}{HTML}{E2CDE4}  
\definecolor{color5}{HTML}{E9D5D4}  
\definecolor{color6}{HTML}{FFFFD1}  

\begin{table*}[h!]\footnotesize
    \centering
    \caption{Summary of selected events across domains. 
    Intervention types are color-coded as 
    \colorbox{color1}{Publicity}, \colorbox{color2}{Announcement}, 
    \colorbox{color4}{Prohibition}, \colorbox{color5}{Response}, \colorbox{color6}{Refutation} and public \colorbox{color3}{Comments}.
    The color scheme matches the shaded areas in Figure~\ref{fig6:attitude} and Figure~\ref{fig10:other_opinion}.}
   
    \label{tab5:case event}
    \renewcommand{\arraystretch}{1.2} 
    \begin{tabular}{c l p{2cm} p{6.5cm} p{5cm}}
    \toprule
    Event ID & Domain & Intervention & Description & Distinctive Features \\ 
    \midrule
    \#2  & Education  &  \colorbox{color1}{Publicity} \& \colorbox{color2}{Announcement} & A school cafeteria was reported to have served spoiled pork.  
         & A traffic surge and two explosive events of source agent intervention \\
    \#7  & Education  & \colorbox{color5}{Response}   & A case of academic dishonesty led to a professor’s dismissal.  
         & A high-profile event with a viewership peak on the 3rd day. \\
    \#14 & Education  & \colorbox{color5}{Response}   & 1,477 freshmen forfeited admission due to late enrollment.  
         & A regular event with views peaking on the 2nd day. \\
    \#31 & Sports     & \colorbox{color2}{Announcement}  & The former head coach of the Chinese national football team was sentenced to 20 years in prison in the first instance.  
         & A highly explosive event with views peaking on the 2nd day. \\
    \#44 & Education  & \colorbox{color1}{Publicity}  & A vocational school student ranked 12th in the Global Mathematics Competition but faced skepticism over group-agents \colorbox{color3}{comments}.
         & A high-profile event with two peak. \\
    \#49 & Economy    & \colorbox{color6}{Refutation} & A fire broke out at a World Financial Center in a certain area.  
         & A regular event with views peaking on the 2nd day. \\
    \#54 & Society    & \colorbox{color4}{Prohibition} & A university student went missing in Myanmar.  
         & A high-profile event with a viewership peak on the 3rd day. \\
    \#55 & Politics   & \colorbox{color4}{Prohibition} & A policy on gradually delaying the statutory retirement age will be implemented next year.  
         & A highly explosive event with a surge of traffic. \\
    \bottomrule
    \end{tabular}
\end{table*}

\section{Opinion Dynamics Formulation}
\label{appB:opinion_dynamics}

\noindent\textbf{Definition 1 (DeGroot Model).} \\
Let $G = (V, E)$ denote a social network, where $V$ represents a population of agents 
and $E$ the set of social connections among them. 
Let $\mathcal{N}(i) = \{\, j \mid i \sim j \in E \,\}$ denote the set of neighbors of agent $i \in V$. 
At each timestep $t$, agent $i$ updates its opinion $x_i(t)$ by averaging the opinions of its neighbors:
\begin{equation*}
f(x(t))_i = \frac{1}{|\mathcal{N}(i)|} \sum_{j \in \mathcal{N}(i)} x_j(t)
\end{equation*}
While the DeGroot model has been highly influential in modeling consensus formation and social learning, 
it always converges to a global consensus (typically the mean of initial opinions) under mild conditions. 
Hence, it is insufficient to capture phenomena such as polarization or radicalization.

\vspace{4pt}
\noindent\textbf{Definition 2 (Hegselmann--Krause Model).} \\
The Hegselmann--Krause (HK) model extends DeGroot by introducing a bounded-confidence mechanism, 
which restricts influence to neighbors with sufficiently similar opinions. 
Formally, let
\begin{equation*}
\mathcal{N}(i,t) = \{\, j \mid i \sim j \land |x_i(t) - x_j(t)| \le \epsilon \,\}
\end{equation*}
denote the set of neighbors of agent $i$ whose opinions lie within a confidence bound $\epsilon$. 
The opinion update is then defined as:
\begin{equation*}
f(x(t))_i = \frac{1}{|\mathcal{N}(i,t)|} \sum_{j \in \mathcal{N}(i,t)} x_j(t)
\end{equation*}
By adjusting $\epsilon$, this model generates diverse social phenomena, ranging from consensus to clustering and polarization.

\vspace{6pt}
\noindent\textbf{Our Parameterized Model.}
To unify these two formulations, we introduce a confidence parameter $\epsilon \in [0,1]$ 
that continuously interpolates between the DeGroot and HK dynamics.
Specifically, given opinions normalized to $[0,1]$, we define the adaptive neighborhood as
\begin{equation*}
\mathcal{N}_\epsilon(i,t) = 
\{\, j \mid i \sim j \text{ and } |x_i(t) - x_j(t)| \le \epsilon \,\},
\end{equation*}
with the following limiting cases:
\begin{itemize}[leftmargin=*]
  \item $\epsilon = 0$: the agent is fully self-isolated ($\mathcal{N}_0(i,t) = \{i\}$), preserving its own opinion;
  \item $\epsilon = 1$: the agent interacts with all others ($\mathcal{N}_1(i,t) = V$), equivalent to the uniform DeGroot update;
  \item $0 < \epsilon < 1$: the agent interacts selectively, forming a local neighborhood consistent with bounded-confidence dynamics.
\end{itemize}
The update rule follows:
\begin{equation*}
x_i(t+1) = 
\frac{1}{|\mathcal{N}_\epsilon(i,t)|}
\sum_{j \in \mathcal{N}_\epsilon(i,t)} x_j(t),
\end{equation*}
which allows $\epsilon$ to serve as a continuous control parameter of \textit{interaction openness}.
The confidence threshold $\epsilon$ introduced in Eq.~\ref{eq:action} 
governs the formation of adaptive neighborhoods $\mathcal{N}_\epsilon(i,t)$ 
that determine the range of social influence during attitude updating.

\noindent\textit{\textbf{Proof.}}  
\vspace{3pt}

\noindent
For $\epsilon = 0$, the bounded-confidence condition 
\[
|x_i(t)-x_j(t)| \le 0
\]
holds only when $x_i(t)=x_j(t)$.  
Hence, $\mathcal{N}_0(i,t)=\{i\}$, and the update becomes
\[
x_i(t+1)=x_i(t),
\]
indicating that each agent remains self-isolated without external influence.  

\vspace{4pt}
\noindent
For $\epsilon = 1$, since all opinions are normalized to $[0,1]$,  
the inequality $|x_i(t)-x_j(t)| \le 1$ holds for all $i,j$.  
Thus $\mathcal{N}_1(i,t)=V$, and the update rule reduces to the uniform DeGroot process:
\[
x_i(t+1)=\frac{1}{N}\sum_{j\in V}x_j(t).
\]

\vspace{4pt}
\noindent
Therefore, the proposed formulation provides a continuous bridge between 
the Hegselmann--Krause and DeGroot models through the single parameter~$\epsilon$.

\section{Experimental Details}
\label{appC:experimental_details}

\begin{figure*}
    \includegraphics[width=\linewidth]{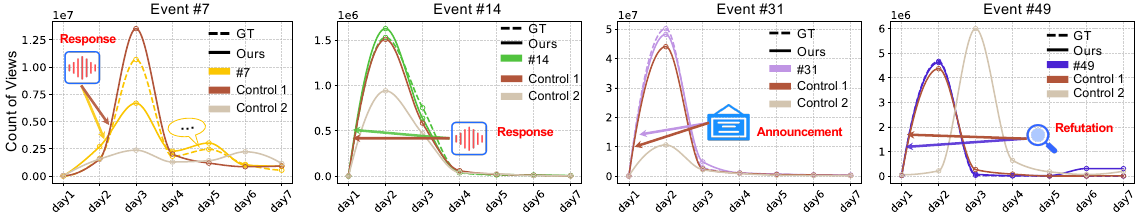}
    \caption{Additional event-level traffic trends under different intervention and control settings, validating the consistency of intervention and timing effects observed in the main results.}
    \label{fig9:other_intervene_control}
\end{figure*}

\begin{figure*}
    \includegraphics[width=\linewidth]{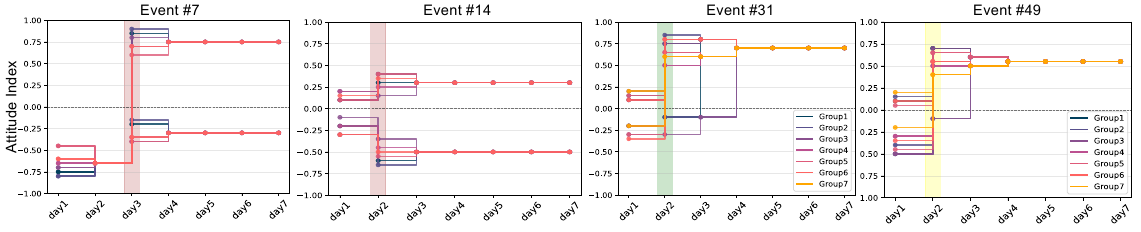}
    \caption{Additional results on opinion dynamics under different interventions. The figure illustrates how \textbf{responses} tend to intensify polarization, while \textbf{announcements} and \textbf{refutations} promote collective consensus.}
    \label{fig10:other_opinion}
\end{figure*}


\subsection{Controlled Intervention}
\label{appB.3:controlled intervention}
As shown in Figure~\ref{fig5:intervene_control} and Figure~\ref{fig9:other_intervene_control}, we provide detailed comparisons across representative controlled experiments to illustrate how different intervention types and timings shape traffic dynamics.

For \textbf{Event \#2}, which involved two sequential interventions—\textit{Publicity} followed by \textit{Announcement}—the experimental setting closely reproduced the real-world traffic pattern. In \textbf{Control 4}, where the second intervention (\textit{Announcement}) was omitted, the second engagement peak disappeared. In \textbf{Control 5}, removing the first intervention (\textit{Publicity}) reduced the initial surge, leading to an overall decline in traffic volume.

For \textbf{Events \#7, \#14, and \#31}, the experimental results were consistent with observed dynamics. In \textbf{Control 1}, restricting communicative responses (\textit{Comments}) caused a moderate decrease in overall traffic. In \textbf{Control 2}, without any intervention (\textit{Response} or \textit{Announcement}), public attention remained significantly lower, underscoring the amplifying effect of intervention-driven exposure.

For \textbf{Event \#44}, the experimental group again matched empirical trends. In the first cycle, \textbf{Control 1} produced lower traffic and no secondary peak, while \textbf{Control 2}, omitting the \textit{Publicity} intervention, resulted in minimal exposure. These findings indicate that early-stage \textit{Publicity} effectively triggered recursive engagement within the discussion network.

For \textbf{Event \#49}, the experimental configuration reproduced the observed engagement trajectory. \textbf{Control 1} showed reduced participation and no secondary peak, whereas \textbf{Control 2}, removing the \textit{Refutation} intervention, yielded higher sustained attention with a delayed peak on day three. Without refutation, discussions continued circulating within the same audience cluster, reinforcing attention but lacking broader diffusion.

For \textbf{Event \#54}, where the experimental condition included a restrictive \textit{Prohibition} intervention, traffic trends aligned with real-world observations. In \textbf{Control 2}, removing the \textit{Prohibition} led to continued escalation of attention and persistently high engagement, but no secondary peak emerged. In \textbf{Control 3}, delaying the \textit{Prohibition} by two days produced a similar pattern with weaker post-intervention variation, suggesting lower responsiveness to late interventions.

For \textbf{Event \#55}, which also featured a \textit{Prohibition} intervention, the experimental outcomes closely mirrored empirical dynamics. In \textbf{Control 2}, the absence of \textit{Prohibition} produced a weak initial response and no subsequent amplification. In \textbf{Control 3}, applying the \textit{Prohibition} earlier (day two) generated a stronger suppressive effect, significantly reducing engagement throughout the cycle and highlighting the timing sensitivity of intervention impacts.

\subsection{Opinion Dynamics Analysis}
\label{appB.4:opinion dynamics}
For \textbf{Event \#2}, the experimental group exhibited two clear attitude shifts among simulated crowd agents. As shown in Figure~\ref{fig6:attitude} (Event~\#2), the first shift followed a negative \textbf{Publicity} intervention, which introduced a framing change that reinforced public dissatisfaction and produced a strong wave of critical sentiment. The second shift occurred after a positive \textbf{Announcement}, which reframed the narrative and prompted a collective move toward supportive evaluations. These two realignments corresponded to distinct engagement peaks, reflecting how early negative framing and subsequent clarification jointly shaped the direction of public sentiment~\cite{ji2026strideedstrategygroundedstepwisereasoning,ji2025hybrid}.

For \textbf{Event \#44}, two major attitude transitions were observed. As shown in Figure~\ref{fig6:attitude} (Event~\#44), an initial \textbf{Publicity} intervention amplified early alignment, shifting public response from mild polarization toward stronger affirmation and reduced skepticism. Later, emerging skeptical comments introduced narrative contestation, triggering a polarity reversal with a higher share of critical responses. The two engagement peaks thus reflect how external publicity and internal discussion jointly drive recursive opinion shifts.

For \textbf{Event \#54}, a restrictive \textbf{Prohibition} intervention limited event-related discussion and weakened the ongoing spread of negative sentiment (see Figure~\ref{fig6:attitude}, Event~\#54). In its aftermath, some subgroups generated compensatory positive narratives, producing mild polarization with overall lower expressive intensity. Although a modest engagement peak emerged, its small magnitude suggests that information-flow constraints prevented large-scale amplification.

For \textbf{Event \#55}, a delayed \textbf{Prohibition} intervention produced the opposite effect (Figure~\ref{fig6:attitude}, Event~\#55). By the time the restriction was introduced, critical narratives had already consolidated in public response. The belated intervention failed to reorganize the dominant stance and instead reinforced existing negativity. Although minor positive reactions appeared briefly, they were quickly absorbed by the prevailing critical tone, illustrating the limited impact of delayed interventions on entrenched attitudes.


As shown in Figure~\ref{fig10:other_opinion}, 
\textbf{Event~\#7} and \textbf{Event~\#14} (responses) exhibit clear polarization patterns, 
where divergent interpretations and emotional amplification drive opinion bifurcation following the initial engagement peaks.  
In contrast, \textbf{Event~\#31} (announcement) and \textbf{Event~\#49} (refutation) display emerging consensus, 
as authoritative communication and factual clarification effectively realign public sentiment toward a shared evaluative stance.  
These observations further substantiate the general trend identified in Figure~\ref{fig7:heatmap}, 
indicating that \textbf{responses} often induce polarization, whereas \textbf{announcements} and \textbf{refutations} tend to consolidate consensus within public response.

\section{Ethical Considerations}
\label{appD:ethical}
IntervenSim is intended as a research framework for intervention-aware social simulation, rather than a decision-making system for real-world governance or platform control. Because it models source-crowd interaction and opinion dynamics under intervention, its outputs could be over-interpreted as prescriptions for how institutions should intervene in public discourse. We do not advocate such use.

(1) \textbf{Abstraction of social heterogeneity.}
Crowd agents improve scalability but necessarily compress within-group diversity in cognition, motivation, and expression~\cite{e1-1,e1-2,e1-3}. As a result, simulated attitude shifts or opinion dynamics should not be treated as faithful proxies for all individuals represented by a group~\cite{e1-4,e1-5}.

(2) \textbf{Dependence on LLM representation fidelity.}
Group agents and their reasoning traces are generated by large language models, whose internal representations may reflect stereotypes, uneven cultural coverage, or biased associations\cite{e2-1,e2-2-acl-2023,e2-3-ACL-2024,dai2026tearscheersbenchmarkingllms}. This may distort the portrayal of certain communities and misrepresent social heterogeneity~\cite{NMI-LLM-study-group-harm-2025large,e2-4}.

(3) \textbf{Misuse risk of intervention modeling.}
Because the framework explicitly models intervention timing, types, and effects on collective response, it could in principle be misused to explore strategies for amplifying, suppressing, or reframing public attention. Our goal is scientific analysis and comparative simulation, not operational influence deployment~\cite{e3-1,e3-2,e4-2}.

(4) \textbf{Simplified platform dynamics.}
The current framework does not model recommendation, ranking, moderation, or cross-platform diffusion. It may therefore omit important sources of amplification, suppression, or feedback in real-world information environments~\cite{e4-1}.

(5) \textbf{Data and privacy considerations.}
Our experiments use event-level social media data and aggregate traffic trajectories rather than individual user modeling. Nevertheless, sensitive events may still contain harmful narratives, so any future release of data, prompts, or code should follow platform terms, privacy constraints, and responsible disclosure practices~\cite{e5-1}.

\section{Prompt Design}
\label{appE:prompt}
We present the main prompts used in our framework, covering relevance reasoning for subgroup generation~\ref{tab6:finegrain_reasoning_prompt}, source-side intervention~\ref{tab7:source_prompt}, crowd deliberation~\ref{tab8:crowd_agent_prompt}, and group-level engagement estimation~\ref{tab9:group_y4_prompt}. 

\tcbset{
    colback=gray!10,    
    colframe=black,     
    width=\textwidth,   
    boxrule=0.5mm,      
    arc=4mm,            
    title=,             
    enhanced,           
}

\begin{table*}[htbp]
\centering
\begin{tcolorbox}[title=]
\textbf{System:} \\
You are a relevance reasoning agent responsible for fine-grained group specialization. \\
Your task is to expand a general population node (e.g., \texttt{students}) into finer subgroups and evaluate their relevance to a given event.

\begin{tabbing}
Event Context: \\
\hspace{1em} \= \textbullet\ \texttt{\{event\_description\}} \\[0.5em]
Coarse Group Node: \\
\hspace{1em} \= \textbullet\ \texttt{\{group\_node\}} \\
\end{tabbing}

\textbf{Instructions:} \\
\textbf{1. Generation:} \\
\hspace{1em} \textbullet\ Generate a list of possible fine-grained subgroups under the provided general group node, grounded in the event content. \\
\hspace{1em} \textbullet\ Subgroups should reflect behavioral or contextual distinctions (e.g., discipline, role, engagement type). \\[0.5em]

\textbf{2. Scoring:} \\
\hspace{1em} \textbullet\ For each generated subgroup, assess how semantically aligned it is with the event content. \\
\hspace{1em} \textbullet\ Use relevance reasoning based on LLM contrastive prompts to assign a score between 0 and 1. \\[0.5em]

\textbf{Output Format (JSON):}
\begin{tabbing}
\hspace{1em} \= \texttt{\char`\{} \\
\hspace{2em} \= "generated\_subgroups": [ \\
\hspace{3em} \= \texttt{\char`\{}"name": "\{subgroup\_name\}", "relevance": \{score\}\texttt{\char`\}}, \\
\hspace{3em} \= ... (multiple subgroup entries) ... \\
\hspace{2em} \= ] \\
\hspace{1em} \= \texttt{\char`\}} 
\end{tabbing}

\textbf{Notes:} \\
• The top-$k$ most relevant subgroups will be used for agent instantiation in subsequent simulation. \\
• Relevance scores should be consistent with the semantic fit of the subgroup to the event context.

\rule{\textwidth}{0.4pt}
\textbf{Example Output:}
\begin{tcolorbox}[colback=white!97!gray, colframe=black!70, arc=2mm, boxrule=0.4pt]
\begin{lstlisting}[basicstyle=\ttfamily\small]
{
  "generated_subgroups": [
    {"name": "master of mathematics", "relevance": 0.91},
    {"name": "bachelor", "relevance": 0.85},
    {"name": "spectator", "relevance": 0.72},
    {"name": "researcher", "relevance": 0.42}
  ]
}
\end{lstlisting}
\end{tcolorbox}

\end{tcolorbox}
\caption{Relevance reasoning prompt for fine-grained subgroup generation and evaluation}
\label{tab6:finegrain_reasoning_prompt}
\end{table*}

\begin{table*}[htbp]
\centering
\begin{tcolorbox}[title=] 
\textbf{System:} \\
You are \{source\_agent\_name\}, a government/institutional/media/opinion-leader actor responsible for managing public information flow. \\
You operate within the social network world: \{world\_description\}, where events evolve through public response and agent interactions.

\begin{tabbing}
perception: \\
\hspace{1em} \= • \= Time: \{day\_n\} \\
\hspace{1em} \= • \= Current Event: \{event\_description\} \\
\hspace{1em} \= • \= Public Opinion: \{opinion\_summary\} \\
\hspace{1em} \= • \= Current Visibility: \{visibility\} \\
Your State: \\
\hspace{1em} \= • \= Intervention History: \{intervention\_history\} \\
\hspace{1em} \= • \= Policy Priority: \{policy\_goal\} \\
\hspace{1em} \= • \= Assessment of Public Risk: \{risk\_assessment\}
\end{tabbing}

\textbf{Reason step-by-step:} \\
• Based on event status, public opinion, and your institutional goal, reason step-by-step to decide which intervention type to apply. \\
• Explain how your decision aligns with minimizing misinformation, guiding public attention, or stabilizing opinion dynamics. \\
• Your reasoning should explicitly connect current event visibility, opinion polarity, and risk evaluation with your chosen intervention strategy.\\

\textbf{Output Format:} The output must follow the JSON structure below:
{
\small
\begin{verbatim}
{
  "reasoning_trace": "Detailed chain-of-thought explanation",
  "selected_intervention": "Prohibition | Refutation | Publicity | Response | Announcement",
  "expected_effect": "Visibility ↓/↑ | Opinion shift | Consensus building | Polarization mitigation"
}
\end{verbatim}
}

\rule{\textwidth}{0.4pt}

\textbf{Intervention Options ($I_1$–$I_5$):} \\
A source agent may select one of five interventions according to current conditions:\\
\textbf{(1) Prohibition} — restrict event visibility to suppress escalation; \\
\textbf{(2) Refutation} — counter misinformation and restore factual accuracy; \\
\textbf{(3) Publicity} — enhance exposure to guide public focus toward official narratives; \\
\textbf{(4) Response} — provide institutional or leader feedback to shape interpretation; and \\
\textbf{(5) Announcement} — issue formal statements that reframe event meaning or provide closure. \\ 
Each intervention aims to modulate information diffusion, emotional salience, or collective interpretation depending on situational needs.

\end{tcolorbox}
\caption{Source-agent prompt for decision-making over five intervention types ($I_1$–$I_5$).}
\label{tab7:source_prompt}
\end{table*}

\begin{table*}[htbp]
\centering
\begin{tcolorbox}[title=] 
{\small
\textbf{System:} \\
You are \{agent\_name\}, a crowd agent representing collective audience opinions. \\
You operate in the social network world: \{world\_description\}, where events evolve through communication and reflection among agents.

\begin{tabularx}{\textwidth}{@{}X X@{}}
\textbf{Perception:} &
\textbf{Your State:} \\[2pt]
• Time: \{day\_n\} &
• Cognitive State: \{previous\_state\} \\
• Current Event: \{event\_description\} &
• Memory Trace: \{memory\_trace\} \\
• Source-side Interventions: \{intervention\_vector\} &
• Interaction Range: \{interaction\_range\} \\
• Aggregated Public Signal: \{mean\_state\} &
• Opinion: \{opinions\} \\
\end{tabularx}

\textbf{Reason step-by-step:} \\
• Considering your cognitive state, memory, and current opinions, reason step-by-step to decide your next deliberative action. \\
• Your reasoning should connect source-side interventions, neighboring opinions, and emotional influence to your action choice. \\
• Explicitly describe how you balance cognition with collective communication when selecting actions.\\
\textbf{Output Format:} The output must follow the JSON structure below:
\begin{verbatim}
{
  "reasoning_trace": "Detailed chain-of-thought explanation",
  "selected_action": "SelectPartner | DiscussOpinion | UpdateOpinion"
}
\end{verbatim}

\rule{\textwidth}{0.4pt}

\textbf{Action 1: SelectPartner} \\
Decide your level of openness to interaction using parameter $\epsilon \in [0,1]$.  
$\epsilon=1$ indicates a fully connected network (open to all agents),  
$\epsilon=0$ means self-isolation, and $0<\epsilon<1$ reflects partial interaction within a local neighborhood.  
Select potential discussion partners according to this openness.

\textbf{Output Format (JSON):}
\begin{verbatim}
{
  "action": "SelectPartner",
  "epsilon": float in [0,1],
}
\end{verbatim}

\rule{\textwidth}{0.4pt}

\textbf{Action 2: DiscussOpinion} \\
Engage in discussions with each selected neighbor individually.  
For every neighboring agent, respond with a message reflecting your current opinion, emotional tone, and interpretation of the event.  
Each response should indicate how you agree, disagree, or remain neutral toward the neighbor’s stance.

\textbf{Output Format (JSON):}
\begin{verbatim}
{
  "action": "DiscussOpinion",
  "responses": [
    {
      "target_agent_id": "agent_id_n",
      "reply_content": "your message or comment",
      "reply_tone": "supportive | neutral | opposing"
    },
    ...
  ]
}
\end{verbatim}

\rule{\textwidth}{0.4pt}

\textbf{Action 3: UpdateOpinion} \\
Update your opinion based on the influence of \textbf{source-side interventions} or \textbf{high-impact public discussions}.  
The opinion score lies within $[-1,1]$, where negative values denote opposition and positive values indicate support.  
Adjust your stance considering group consensus, emotional contagion, and authoritative signals.

\textbf{Output Format (JSON):}
\begin{verbatim}
{
  "action": "UpdateOpinion",
  "updated_opinion": float in [-1,1],
  "update_reason": "source_intervention / group_discussion / self_reflection",
}
\end{verbatim}

}
\end{tcolorbox}
\caption{Crowd-agent prompt for deliberative reasoning with three actions: interaction selection, discussion, and opinion updating.}
\label{tab8:crowd_agent_prompt}
\end{table*}

\begin{table*}[htbp]
\centering
\begin{tcolorbox}[title=] 
\textbf{System:} \\
You are \{agent\_name\}, a crowd agent responsible for generating group-level engagement indicators 
based on collective opinion and social context. 
You operate in the social network world: \{world\_description\}, 
where source-side interventions and crowd deliberations jointly influence information diffusion.

\vspace{4pt}
\textbf{Perception:}\\
• Time: \{day\_n\} \\
• Current Event: \{event\_description\} \\
• Source-side Interventions: \{intervention\_vector\} \\
• Aggregated Public Signal: \{mean\_state\} \\
\textbf{Your State:}\\
• Cognitive State: \{previous\_state\} \\
• Memory Trace: \{memory\_trace\} \\
• Interaction Range: \{interaction\_range\} \\
• Opinion: \{opinions\} \\

\vspace{4pt}

\textbf{Reason step-by-step:} \\
• Analyze how collective opinion, interaction range, and source-side influence shape engagement behavior. \\
• Stronger positive opinions and wider interaction ranges generally correspond to higher engagement levels. \\
• Incorporate contextual cues from interventions and aggregated public signals when generating results.

\textbf{Output Format:} The output must follow the JSON structure below:
\begin{verbatim}
{
  "reasoning_trace": 
    "Reasoning on how opinion and group features jointly affect engagement.",
  "engagement_vector": {
    "views": int,
    "likes": int,
    "comments": int,
    "shares": int
  },
  "generation_reason": 
    "Derived from opinion intensity, group characteristics, and external influence."
}
\end{verbatim}

\end{tcolorbox}
\caption{Prompt for generating group-level engagement indicators ($y_4$) conditioned on opinion, interaction, and source-side influence.}
\label{tab9:group_y4_prompt}
\end{table*}

\end{document}